# Sharp steps in magnetization, magnetoresistance and magnetostriction in $Pr_{0.6}Sr_{0.4}Co_{1-y}Ga_yO_3$

A. Chanda and R. Mahendiran[1]

[1]Department of Physics, National University of Singapore, 2 Science Drive 3,

Singapore -117551, Republic of Singapore

**Abstract**

We report the effect of Ga substitution on magnetization, magnetoresistance, and magnetostriction in polycrystalline $Pr_{0.6}Sr_{0.4}Co_{1-y}Ga_yO_3$ ($y$=0.0-0.3) samples. Upon substitution of the non-magnetic $Ga^{3+}$ cation for $Co^{3+}$, the low temperature ground state transforms from ferromagnetic metallic for $y = 0$ to cluster glass semiconductor for $y = 0.2$. Magnetoresistance at 7T is negative and its magnitude increases from 2% for $y = 0$ to 30% for $y = 0.3$ at 10 K. On the other hand, magnetostriction at 10 K is positive and its value decreases with increasing $y$. Interestingly, the field dependent magnetization, magnetoresistance and magnetostriction for $y \geq 0.2$ and at $T \leq 3$ K show reversible abrupt steps for both positive and negative magnetic field whereas all these quantities vary smoothly with the magnetic field above 4 K. Such steps in all three distinct physical quantities were never reported earlier in perovskite cobaltites and they differ from observations made in manganites and intermetallic alloys. It is suggested that field-induced avalanche flipping of ferromagnetic clusters could be the origin of observed steps in all these three quantities.

---

[1] Email: phyrm@nus.edu.sg

The perovskite cobaltites $R$CoO$_3$ [$R$=rare-earth] are nonmagnetic insulators at low temperatures but become ferromagnetic metals upon introduction of holes through partial substitution of divalent Sr$^{2+}$ cation for $R^{3+}$ cation when $x \geq 0.2$.[1] Although both hole-doped manganites and cobaltites are ferromagnetic metals for doping levels $x$=0.2-0.5, magnetoresistance (MR) around the ferromagnetic Curie temperature $T_C$ is much larger in manganites than cobaltites, $e.g.$, $MR \approx 40\%$ for $\mu_0H$=6T in La$_{0.6}$Sr$_{0.4}$MnO$_3$[2] whereas it is only 6% in La$_{0.6}$Sr$_{0.4}$CoO$_3$.[3,4] In contrast, lower compositions ($x$=0.12-0.18) of La$_{1-x}$Sr$_x$CoO$_3$ series show giant $MR \approx 70\%$ for $\mu_0H$=6T at $T$=5K.[3] Such giant MR was attributed to spin dependent tunneling between ferromagnetic clusters which are randomly dispersed in non-magnetic matrix.[5] Exploration of magnetotransport in hole-doped rare-earth cobaltites other than $R$=La did not evoke much enthusiasm among researchers because the magnitude of $MR$ at $T=T_C$ decreases with decreasing size of $R^{3+}$ or divalent alkaline earth ions.[6,7,8,9] An intriguing property of cobaltites is the existence of giant anisotropic magnetostriction in La$_{1-x}$Sr$_x$CoO$_3$ series ($x$=0.3-0.5) attributed to magnetic field-induced spin-state transition of Co ions at low temperatures,[10,11,12] however, magnetostriction in other rare-earth cobaltites is scarcely reported until now.[13,14]

In this letter, we report the influence of Co-site substitution on the magnetization, magnetoresistance and magnetostriction in Pr$_{0.6}$Sr$_{0.4}$Co$_{1-y}$Ga$_y$O$_3$ ($y$=0-0.3). The substitution of non-magnetic Ga$^{3+}$ ions is expected to dilute ferromagnetic interactions among Co$^{3+}$ and Co$^{4+}$ ions. An earlier work on magnetic and magnetotransport properties in La$_{0.7}$Sr$_{0.3}$Co$_{1-y}$Ga$_y$O$_3$ showed transformation of the long range ferromagnetic ground state into cluster spin glass state for $y$>0.2.[15]

Here, we show that the magnitude of magnetoresistance dramatically increases with increasing Ga content and show qualitatively different behaviors above and below 4K. Magnetization, magnetoresistance and magnetostriction vary smoothly with increasing magnetic field strength for $T>3K$ whereas abrupt steps occur in all these three quantities at $T\leq 3K$ for $y\geq 0.2$ in $Pr_{0.6}Sr_{0.4}Co_{1-y}Ga_yO_3$. Simultaneous observation of these steps in magnetization, magnetoresistance and magnetostriction in cobaltites has never been reported earlier.

Polycrystalline samples of $Pr_{0.6}Sr_{0.4}Co_{1-y}Ga_yO_3$ ($y=0-0.3$) were prepared using standard solid state reaction method. Electrical resistivity and magnetization were measured using a Physical Property Measuring System (PPMS) with a vibrating sample magnetometer option. From the field dependence of resistivity measured at a constant temperature, magnetoresistance is estimated using the definition $MR(\%)=[\rho(H)-\rho(H=0)]/\rho(H=0)\times 100$. Longitudinal magnetostriction, $\lambda_{//}$ (defined as $\lambda_{//}=[L(H)-L(H=0)]/L(H=0)$, where, $L(H=0)$ is the sample length when $H=0$) as a function of magnetic field was measured using a miniature capacitance dilatometer probe attached to the PPMS.

Fig. 1(a) shows $M(T)$ of the series $Pr_{0.6}Sr_{0.4}Co_{1-y}Ga_yO_3$ ($y=0-0.3$) measured in zero field cooled (ZFC) and field-cooled (FC) modes for $\mu_0H=0.1T$. The parent sample $y=0$ undergoes a paramagnetic (PM)-ferromagnetic (FM) transition at $T_C=214K$ and shows a step-like anomaly at $T_S=69K$, which is triggered by orthorhombic-tetragonal structural transition.[16] $T_C$ decreases upon Ga substitution ($T_C = 167, 120, 75$ K for $y=0.1, 0.2$ and $0.3$) whereas the structural transition related anomaly decreases to 56K in $y=0.1$ and disappears for $y\geq 0.2$. The bifurcation between ZFC and FC-$M(T)$ starts much below $T_C$ and the difference increases with $y$. While ZFC-$M(T)$ continues to decrease down to the lowest temperature for $y\leq 0.1$, it tends to be flat at low temperatures for $y\geq 0.2$, which indicates a change in the ground state. The bifurcation in y=0 sample is due to inherent magnetocrystalline anisotropy[16] and it vanishes for $\mu_0H>0.15$T. While resistivity $\rho(T)$ shows metallic behavior above and below $T_C$ in y=0 and 0.1, $\rho(T)$ is semiconducting like in y = 0.1 and 0.2. (see inset of Fig. 1(a)). $\rho$ at 10K increases by 5 orders of magnitude from 620 μΩcm for y= 0 to 42 Ω cm for y = 0.3 reflecting changes in the ground state. Fig. 1(b) shows ZFC and FC-$M(T)$ for y=0.3 recorded under different magnetic fields. The cusp in ZFC-$M(T)$ at $T=T_f$ shifts towards low temperatures with increasing magnetic field strength but it is present even at 2T albeit for a small temperature range. Long range ferromagnetism is most likely absent for y≥0.2, instead the ground state is likely to be composed of unlinked ferromagnetic clusters. Magnetization within these clusters are frozen in random directions as dictated by local anisotropy. Fitting of the $T_f(H)$ data for $\mu_0H \leq 1$T (see inset of Fig. 1(b)) with the Almeida–Thouless (AT) model (governed by the expression: $T_f(H) = T_f(H) - a.H^n$)[17] yields n=0.24 indicating cluster-glass (CG) magnetic ground state for y≥0.2.

The *M-H* loops at 10K exhibit hysteresis (see Fig. 2(a)) and coercive field increases with increasing Ga-content whereas the magnetization value at 7T ($M_{7T}$) decreases. Figs. 2(b) and (c) show the *M-H* loops of y=0.3 for $T \leq 3$K and $T \geq 4$K, respectively. While *M* varies smoothly with *H* for $T \geq 5$ K, *M(H)* at 2K shows a step-like decrease at $H_S$ = -2.2T while decreasing the field from +7T to –7T. A symmetrical curve is traced in the backward field sweep $\mu_0H$= -7T→+7T. In the course of this step-like decrease, *M* reduces by $\Delta M$=0.48μ$_B$. The step occurs at a slightly higher field $H_S$= ±2.4T at 3K. Interestingly, the step in *M* disappears completely at 4K (see inset of Fig. 2(c)) suggesting that the step-like behavior is extremely sensitive to the temperature variation.

Fig. 3(a) displays the $H$-dependence of longitudinal magnetostriction ($\lambda_{//}$) at 10K for $y$=0-0.3. The sign of magnetostriction is positive in all four samples, *i.e.*, $\lambda_{//}$ increases with $H$ and it does not saturate at 7T. The value of $\lambda_{//}$ is the largest for $y$=0 and $\lambda_{//}$ shows a pronounced hysteresis in the whole field range. This large hysteresis is possibly due to field-induced structural transition.[16] The magnitude of $\lambda_{//}$ at 7T and the hysteresis in $\lambda_{//}(H)$ curve decrease with increasing $y$. Fig. 3(b) shows $\lambda_{//}(H)$ isotherms at $T\leq$3K for $y$=0.3. Unlike at 10K, $\lambda_{//}(H)$ at $T\leq$3K increases abruptly at $H=\pm H_S$, where $M(H)$ also showed steps. The inset shows $\lambda_{//}(H)$ curves in an enlarged scale. It is evident from the Fig. 3(c) that the steps are absent for $T\geq$5K.

Fig. 4(a) shows $H$-dependence of magnetoresistance, $MR(H)$ at 10K for $y$=0-0.3. $MR$ is negative in all the samples and the absolute value of magnetoresistance at 7T increases substantially with increasing Ga-content. Interestingly, $MR(H)$ for $y\geq$0.2 shows butterfly-like hysteresis upon cycling the $H$-direction. Fig. 4(b) shows $MR(H)$ for $y$=0.3 at $T\leq$4K. The $MR$ at 2K shows an abrupt decrease (increase) at $H_S$= -2.2T (+2.2T) while decreasing (increasing) the field from +7T (-7T) to -7T (+7T). This is consistent with the magnetization step ($M$-step) observed at 2K. $H_S$ increases to $\pm$2.4T when $T$=3K. In the course of step, $MR$ changes by $\approx$20% for $T\leq$3K. However, the step vanishes for $T\geq$4K as can be seen in the Figs. 4(b) and (c). $MR$ decreases rapidly for $T>$20K. Main panel of Fig. 4(d) shows $MR$ at 2K measured under different $H$-sweep rates: 25, 50, 75 and 90 Oe/s. The position of the step moves only slightly to lower fields with increasing sweep rate (see inset of Fig. 4(d)). We have also compared the $MR$ isotherms recorded after zero field-cooling and field-cooling under +7T from 350 to 2K in the Figs. 4(e) and (f) for $y$=0.2 and 0.3, respectively. The steps in $MR$ in the field-cooled mode occurs at the same field as the zero field-cooled mode for both the samples although there is an exchange bias effect at low fields.

In Fig. 5, we combine the $H$-dependence of magnetization (top), magnetostriction (middle) and magnetoresistance (bottom) at 2K for $y$=0.3 ($y$=0.2) in the left (right) column. We have also shown the initial $H$-sweep after cooling the sample to 2K in zero field. For $y$=0.3, the step-like increase in magnetization at $H_S$ =2.2T during the $H$-sweep (-7T→+7T) is accompanied by an abrupt increase of the magnetostriction but an abrupt decrease of $MR$. However, the steps are less prominent in the initial $H$-sweep (0→7T) although a slope change is present in each of the three quantities. Interestingly, $M$ (also $MR$) in the initial $H$-sweep extends beyond the envelope traced by subsequent $H$-sweeps, which is uncommon in a long-range ferromagnet. For $y$=0.2, $M(H)$ shows two steps at $H_{S1}$=±1.2T and $H_{S2}$=±1.8T. It is evident that $\Delta M$ is higher in magnitude at $H_{S1}$ than at $H_{S2}$. Magnetostriction shows a prominent step at $H_{S1}$ whereas the step at $H_{S2}$ is too weak to detect. On the other hand, $MR$ shows a prominent step at $\pm H_{S1}$ followed by slope change at $\pm H_{S2}$.

Multiple steps in $M(H)$ and $MR(H)$ were first reported in Mn-site doped manganites $Pr_{0.5}Ca_{0.5}Mn_{0.95}M_{0.05}O_3$ [$M$=Co, Ga],[18,19] but later also found in A-site doped manganites *e.g.*, $(Eu_{0.4}La_{0.1})(Sr_{0.4}Ca_{0.1})MnO_3$,[20] $Pr_{1-x}(Ca_{1-y}Sr_y)_xMnO_3$,[21,22] $(La_{0.225}Pr_{0.4})Ca_{0.375}MnO_3$,[23,24] $(La,Ba)_{0.33}Ca_{0.67}MnO_3$,[25] $Sm_{1-x}Sr_xMnO_3$($x$=0.45, 0.5),[26] $Sm_{0.5}(Ca_{0.25}Sr_{0.25})MnO_3$,[27] etc. These steps are seen in polycrystals, single crystals and thin films suggesting intrinsic origin of these steps. Phase separation is crucial to understand the steps in $M$ and $MR$ in these manganites. Ferromagnetic (FM) metallic and charge-ordered (CO)-antiferromagnetic (AFM) insulating phases coexist in zero field and these phases have different lattice parameters or crystallographic structures. For $T\leq 5K$, reduced thermal fluctuations make it difficult for the more distorted CO-AFM phase to be transformed into the less distorted FM phase. When the applied field overcomes the elastic energy barrier, a burst-like growth of the FM phase occurs at the expense of CO-AFM phase and the net magnetization increases along with release of interfacial strain giving rise to multiple magnetization steps. This is the martensitic-like transformation under a magnetic field. In Ti-doped $Pr_{0.5}Ca_{0.5}MnO_3$, ferromagnetic phase is absent in zero field but CE and pseudo CE-type AFM orderings coexist in differently distorted domains.[28] Field-induced transformation of CE-AFM to FM phase causes abrupt increase of magnetization. *M*-steps below 5K were also reported in intermetallics *e.g.*, $Gd_5Ge_4$,[29] Ru-doped $CeFe_2$,[30] $Nd_5Ge_3$,[31] $La_{1-x}Ce_xFe_{12}B_6$.[32] However, all the above-mentioned systems are antiferromagnets without doping and ferromagnetism is induced by doping or by application of a high magnetic field. In the above-mentioned compounds, *M*-steps appear only in the virgin cycle but absent in subsequent field sweeps. Besides, they show temperature-driven first-order transitions under a magnetic field. Multiple steps in *M* occur for both positive and negative fields in double perovskite $Y_2CoMnO_6$.[33] *M*-steps for all the above-mentioned samples move to lower fields with increasing *H*-sweep rate but move to higher *H*-values upon field-cooling. Although *M* and *MR*-steps are weakly dependent on *H*-sweep rate in our samples, the steps are unaffected by field-cooling. So, martensite-like scenario is unlikely the origin of *M*-steps in our samples.

Replacing $Co^{3+}$ ions with non-magnetic $Ga^{3+}$ ions disrupts ferromagnetic interaction between $Co^{3+}$ and $Co^{4+}$ ions. Magnetization data suggests that the magnetic ground state for $y \geq 0.2$ can be viewed as a cluster-glass wherein ferromagnetic clusters (containing mostly $Co^{3+}$-$Co^{4+}$ ions) are randomly frozen in paramagnetic semiconducting matrix (containing $Ga^{3+}$-$Co^{4+}$). The absence of steps in $y \leq 0.1$ suggests that the steps are related to the cluster-glass ground state in $y \geq 0.2$. In zero field, magnetization of each FM cluster points in the direction of local anisotropy axis but anisotropy axes are random among FM clusters. As a result, *M* is very small at zero field. With increasing magnetic field, the FM clusters tend to orient at an angle to the magnetic field as favored by competition between Zeeman and anisotropy energies. Magnetization of majority of these clusters are oriented closer to the direction of the applied field when the field is very high (7T). As the field is reduced and increased in reverse direction, magnetization of a FM cluster is flipped along the negative field direction when the anisotropy energy is balanced by the magnetostatic energy. Energy released during flipping of a FM cluster propagate rapidly[23,24] and leads to avalanche flipping of other clusters which results in magnetization step(s) as seen in polycrystalline CeNi$_{1-x}$Cu$_x$,[34] layered Sr$_2$CoO$_4$[35]. Magnetoresistance is positive in the field regime between coercive fields as the magnetic moments of clusters are randomly oriented near zero field but shows abrupt decrease when moments of these clusters flip in an avalanche manner. Since these clusters are suddenly aligned along the field direction, there is a sudden increase in the sample length parallel to the field direction which causes abrupt increase in $\lambda_{//}(H)$ as shown in the schematic diagram in the inset of Fig. 3(c)).

**Acknowledgement:** R.M thanks the Ministry of Education, Singapore (grant numbers. R144-000-381-112 and R144-000-404-114)

**Data availability statement**: Authors willing to share the data upon request.

**List of figure captions**

**Fig. 1(a)** Temperature dependence of zero field-cooled (ZFC) and field-cooled (FC) magnetization $M(T)$ of $Pr_{0.6}Sr_{0.4}Co_{1-y}Ga_yO_3$ series ($y=0-0.3$) under $H=0.1T$. Inset shows $\rho(T)$ for $y=0-0.3$ at $H=0$. **(b)** ZFC and FC $M(T)$ of $y=0.3$ under different fields and inset shows Almeida–Thouless fit of the cluster-freezing temperature ($T_f$) vs. $H$ curve.

**Fig. 2(a)** $M$-$H$ loop for $y=0-0.3$ at 10K. $M$-$H$ for $y=0.3$ at **(b)** $T\leq 3K$ and **(c)** $T\geq 5K$, inset of **(c)** shows $M$-$H$ at 4K.

**Fig. 3(a)** Magnetic field dependence of magnetostriction, $\lambda_{//}$ at 10K, $\lambda_{//}$ of $y=0.3$ at **(b)** $T\leq 3K$ (inset shows enlarged view) and **(c)** $T\geq 5K$, inset of **(c)** schematic diagram demonstrating increase of sample length under magnetic field.

**Fig. 4(a)** Magnetic field dependence of magnetoresistance, $MR(H)$ of $y=0-0.3$ at 10K, $MR(H)$ of $y=0.3$ at **(b)** $T\leq 4K$ and **(c)** $T\geq 5K$, **(d)** MR at 2K under different field sweep rates for $y=0.3$, $\rho(H)$ at 2K measured in zero field cooled and field cooled modes for **(e)** $y=0.2$ and **(f)** $y=0.3$.

**Fig. 5(a)** Magnetization **(b)** magnetostriction and **(c)** magnetoresistance at 2K for $y=0.3$ as a function of magnetic field. The right panel shows similar quantities for $y=0.2$.

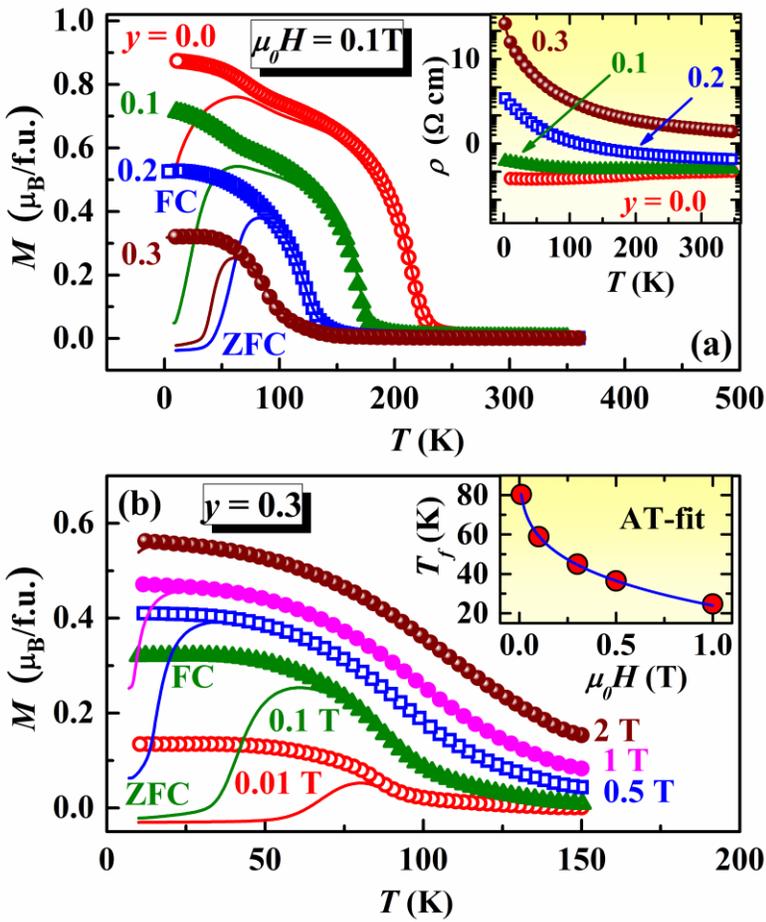

**Fig. 1. A. Chanda et al.**

**Fig. 1(a)** Temperature dependence of zero field-cooled (ZFC) and field-cooled (FC) magnetization $M(T)$ of $Pr_{0.6}Sr_{0.4}Co_{1-y}Ga_yO_3$ series ($y=0-0.3$) under $H=0.1T$. Inset shows $\rho(T)$ for $y=0-0.3$ at $H=0$. **(b)** ZFC and FC $M(T)$ of $y=0.3$ under different fields and inset shows Almeida–Thouless fit of the cluster-freezing temperature ($T_f$) vs. $H$ curve.

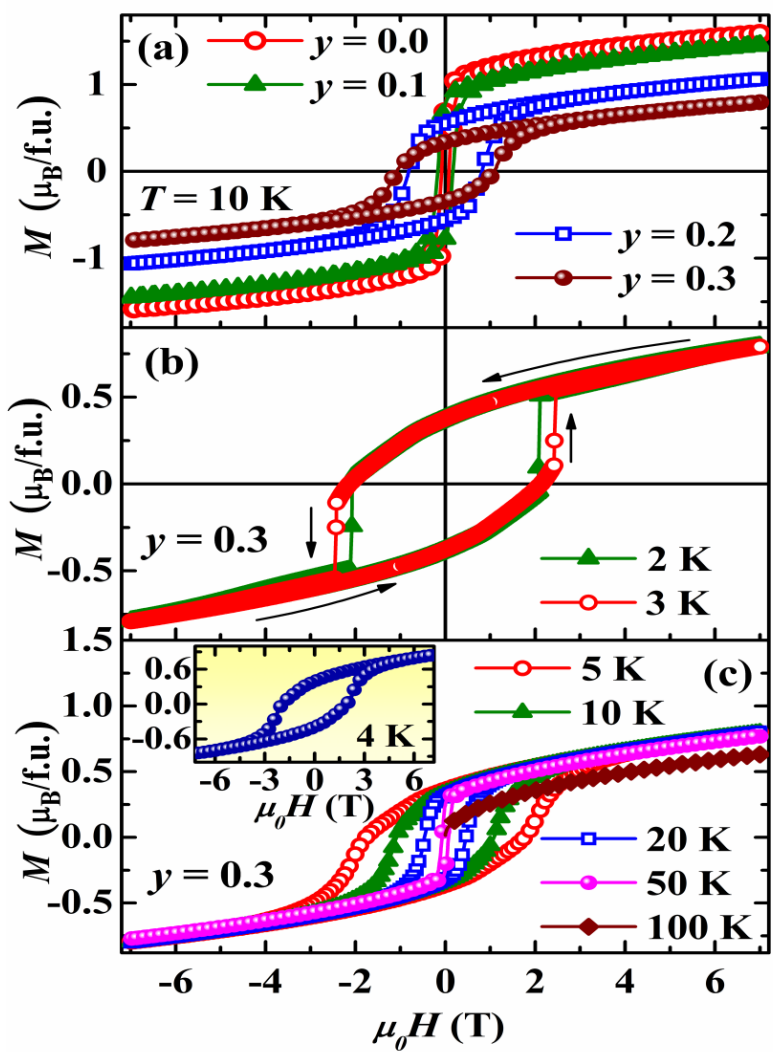

**Fig. 2. A. Chanda et al.**

**Fig. 2(a)** *M-H* loop for *y*=0-0.3 at 10K. *M-H* for *y*=0.3 at (**b**) *T*≤3K and (**c**) *T*≥5K, inset of (**c**) shows *M-H* at 4K

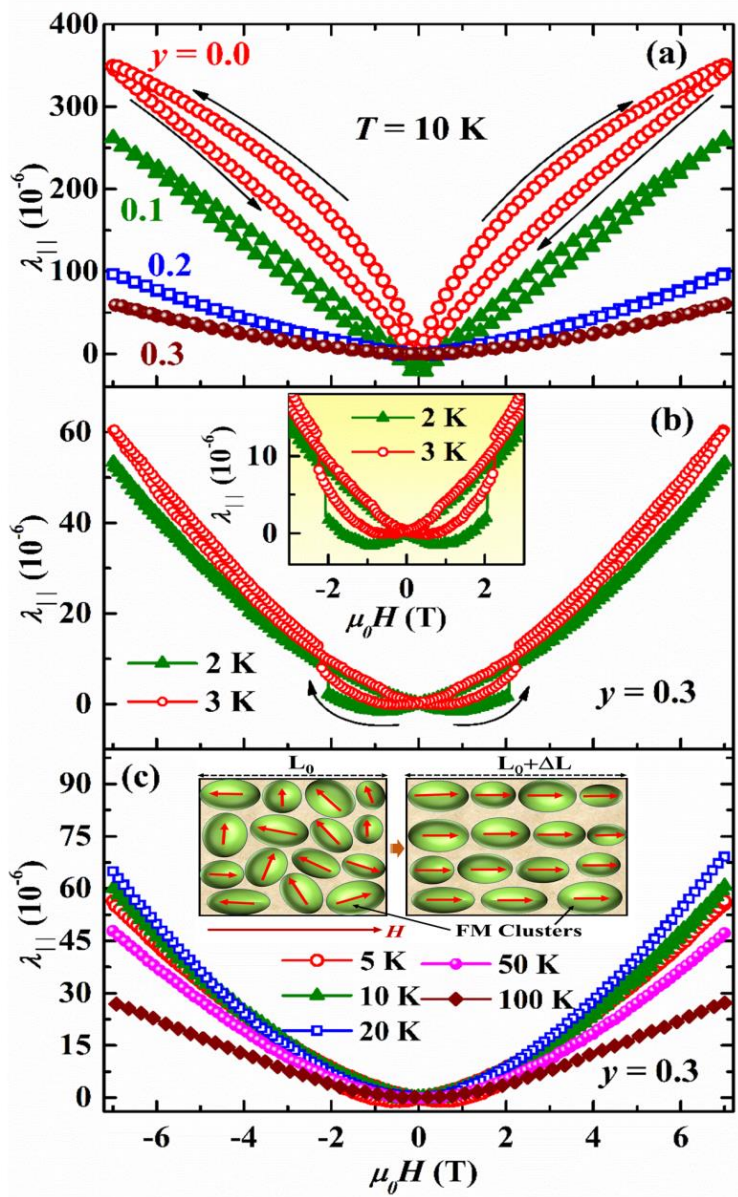

**Fig. 3(a)** Magnetic field dependence of magnetostriction, $\lambda_{//}$ at 10K, $\lambda_{//}$ of $y=0.3$ at **(b)** $T\leq3$K (inset shows enlarged view) and **(c)** $T\geq5$K, inset of **(c)** schematic diagram demonstrating increase of sample length under magnetic field.

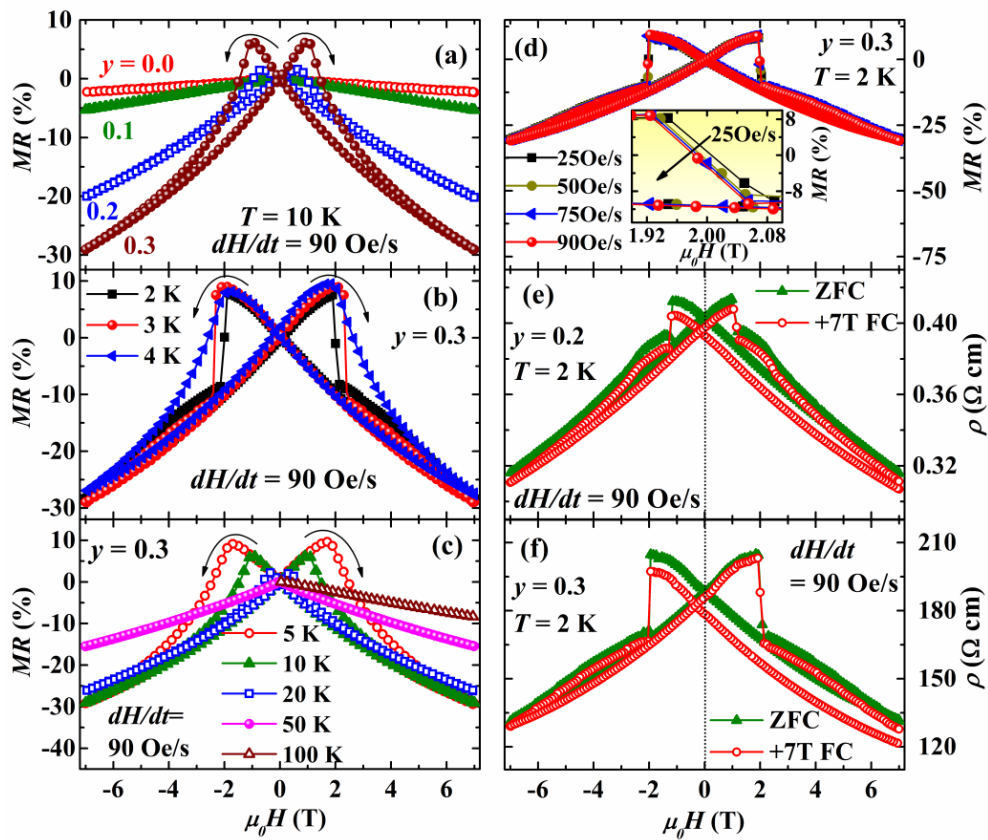

**Fig. 4(a)** Magnetic field dependence of magnetoresistance, *MR(H)* of *y*=0-0.3 at 10K, *MR(H)* of *y*=0.3 at (**b**) *T*≤4K and (**c**) *T*≥5K, (**d**) MR at 2K under different field sweep rates for *y*=0.3, $\rho(H)$ at 2K measured in zero field cooled and field cooled modes for (**e**) *y*=0.2 and (**f**) *y*=0.3.
16

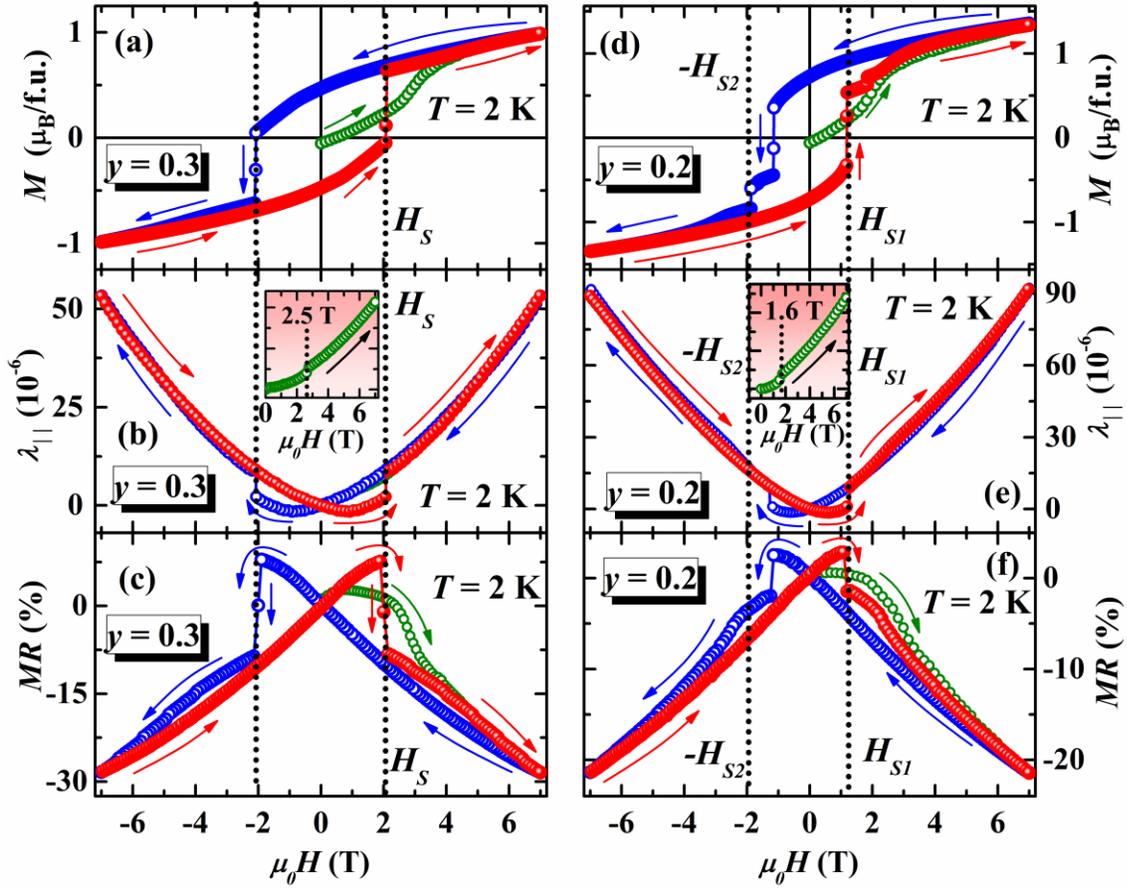

Fig. 5. A. Chanda et al.

**Fig. 5(a)** Magnetization **(b)** magnetostriction and **(c)** magnetoresistance at 2K for *y*=0.3 as a function of magnetic field. The right panel shows similar quantities for *y*=0.2.